\documentclass[12pt]{iopart}
\usepackage{graphicx}
\usepackage{here}

\usepackage{wrapfig}
\usepackage{stackrel}
\usepackage{amssymb,accents,latexsym}

\usepackage{caption}

\newcommand{\be}{\begin{equation}}
\newcommand{\ee}{\end{equation}}
\newcommand{\bea}{\begin{eqnarray}}
\newcommand{\eea}{\end{eqnarray}}

\newcommand{\GW}{\textrm{\scriptsize GW}}

\newcommand{\TTeV}{\textrm{TeV}}
\newcommand{\GeV}{\textrm{GeV}}
\newcommand{\KK}{\textrm{\tiny KK}}
\newcommand{\eff}{\textrm{\scriptsize eff}}
\newcommand{\env}{\textrm{\scriptsize env}}
\newcommand{\sw}{\textrm{\scriptsize sw}}
\newcommand{\rad}{\textrm{\scriptsize rad}}
\newcommand{\m}{\textrm{\scriptsize m}}
\newcommand{\BH}{\textrm{\scriptsize BH}}

\begin{document}

\title[Radion dynamics, heavy Kaluza-Klein resonances and gravitational waves]{Radion dynamics, heavy Kaluza-Klein resonances and gravitational waves}

\author{Eugenio~Meg\'{\i}as$^{1}$, Germano~Nardini$^{2}$ and Mariano~Quir\'os$^{3}$}
\address{
$^{1}$ Departamento de F\'{\i}sica At\'omica, Molecular y Nuclear and Instituto Carlos I de F\'{\i}sica Te\'orica y Computacional, Universidad de Granada, \\ 
Avenida de Fuente Nueva s/n, 18071 Granada, Spain \\

$^{2}$ Faculty of Science and Technology, University of Stavanger, 4036 Stavanger, Norway \\

$^{3}$ Institut de F\'{\i}sica d'Altes Energies (IFAE), Barcelona Institute of  Science and Technology (BIST), Campus UAB, 08193 Bellaterra, Barcelona, Spain
}
\ead{emegias@ugr.es, germano.nardini@uis.no, quiros@ifae.es}

\vspace{10pt}
%\begin{indented}
%\item[]August 2017
%\end{indented}

\begin{abstract}
We study the confinement/deconfinement phase transition of the radion field in a warped model with a polynomial bulk potential. The backreaction of the radion on the metric is taken into account by using the superpotential formalism, while the radion effective potential is obtained from a novel formulation which can incorporate the backreaction. The phase transition leads to a stochastic gravitational wave background that depends on the energy scale of the first Kaluza-Klein resonance, $m_\KK$.  This work completes previous studies in the following aspects: \textit{i)} we detail the evaluation of the radion spectrum; \textit{ii)} we report on the mismatches between the thick wall approximation and the numerical bounce solution; \textit{iii)} we include a suppression factor in the spectrum of sound waves accounting for their finite lifetime; and, \textit{iv)} we update the bound on $m_\KK$ in view of the O3 LIGO and Virgo data. We find that the forthcoming  gravitational wave interferometers can probe scenarios where $m_\KK \lesssim 10^9 \, \TTeV$, while the O3-run bounds rule out warped models with   $10^4 \TTeV \lesssim m_\KK \lesssim 10^7 \TTeV$ exhibiting an extremely strong confinement/deconfinement phase transition.
\end{abstract}

\noindent{\it Keywords}: Physics beyond the Standard Model, extra dimensions,  radion dynamics, electroweak phase transition, gravitational waves.
%\maketitle

\section{Introduction}
\label{sec:introduction}

The Standard Model (SM) of particle physics is unable to explain experimental observations like the baryon asymmetry of the universe and the existence of dark matter, and suffers from theoretical drawbacks, like the electroweak (EW) hierarchy problem, which is understood as a strong sensitivity to high scale physics. These issues have motivated the study of beyond the SM (BSM) scenarios where new physics is present. In particular, the Randall-Sundrum (RS) model was proposed in 1999~\cite{Randall:1999ee} as a way to solve the hierarchy problem between the Planck scale $M_{P} \sim 10^{15} \, \TTeV$ and the EW scale $\rho \sim \TTeV$ by means of a warped extra dimension in Anti-de Sitter (AdS) space. This extra dimension determines a spectrum of heavy Kaluza-Klein (KK) resonances, where the  mass of the first resonance is $m_\KK \sim \rho$. In addition, the model contains a light state, the radion, which is typically the lightest BSM state, with a dynamics dictated by an effective potential. 

The LHC data are putting severe lower limits on the masses of the first KK resonances~\cite{Sirunyan:2018ryr,Aaboud:2019roo}. These limits point toward the possibility that nature might have chosen values of $\rho\gg 1$\,TeV. In such a case  the EW scale would require some amount of tuning, but the warp factor would still naturally explain the hierarchy between the Planck scale and the value of $\rho$. At the same time the KK excitations would easily escape the LHC searches, and so their discovery would rely on future more energetic colliders.

In the present work we investigate the warped models in the large $\rho>1$ TeV regime. To compute the radion effective potential in the presence of large backreactions, we use a recent development of the so called ``superpotential procedure''. In the resulting potential, the radion undergoes a first order phase transition during which it acquires a vacuum expectation value~\cite{Creminelli:2001th, Randall:2006py, Nardini:2007me}. This phase transition, representing a confinement/deconfinement transition, generates a stochastic gravitational wave background (SGWB)~\cite{Caprini:2015zlo} detectable at the present and future gravitational wave interferometers. Thanks to this signature, the warped scenario with KK modes outside the reach of the LHC of future circular colliders, can still be tested.

\section{The Randall-Sundrum model}
\label{sec:RS_model}

In this section we introduce the main properties of the RS model, and outline the objectives that will be pursued in the rest of the paper. The RS model is based on a 5D space with line element~\cite{Randall:1999ee}
\begin{equation}
ds^2 = g_{MN}dx^M dx^N \equiv  e^{-2A(r)} \eta_{\mu\nu} dx^\mu dx^\nu - dr^2 \,, \qquad A(r) = kr \,, \label{eq:RS_metric}
\end{equation}
and two branes, located at different positions along the coordinate $r$ of the extra dimension. The UV brane, at $r = r_0=0$, provides the Planck scale, while the IR brane, at $r=r_1$, sets the TeV (or multi-TeV) scale $\rho$. The hierarchy between these scales is driven by the warped space factor
\begin{equation}
\rho = e^{-k r_1} M_{P}  \qquad \mathrm{with} \qquad k r_1 \sim 35 \,.   \label{eq:hierarchyRS}
\end{equation}
Fields that are mainly localized toward the IR brane are composite; this is the case of the Higgs, heavy fermions and KK modes. On the contrary, fields that are localized toward the UV brane are elementary, as e.g.~light fermions. Other fields, like the zero mode gauge bosons, have a flat profile  in the extra dimension.

In the RS model the brane distance has to be stabilized. This can be achieved by a bulk scalar field~$\phi$ that \emph{softly} breaks conformal invariance and has some bulk and brane potentials fixing its vacuum expectation value~\cite{Goldberger:1999wh}. Due to this gentle breaking, the degree of freedom regulating the brane distance  tends to yield a {\it light state}. The corresponding field, dubbed as radion or dilaton depending on the (5D or 4D) context, is coupled to the SM and exhibits an interesting Higgs-like phenomenology~\cite{Csaki:2000zn}. Nevertheless, the RS model has problems when confronting to EW precision measurements. In particular, it fails to describe oblique observables $S,T,U$, as it leads to too large values of them. Several ideas  have been proposed to circumvent the issue: 
\begin{itemize}
\item Considering a large backreaction on the metric such as to create a singularity. This is achieved by using a deformed metric in the IR~\cite{Cabrer:2011fb,Megias:2018sxv}.
\item Assuming custodial symmetry in the bulk conserved in the IR~\cite{Carena:2018cow}.
\item Moving the KK resonance masses above the TeV scale, e.g.~$m_{\KK}\gtrsim 10$ TeV~\cite{Megias:2020vek}.
\end{itemize}

In this paper we explore the third possibility. It has several implications, in particular the model stops aiming for being a fully naturally solution of the ``whole'' hierarchy problem -- a little hierarchy problem remains --, although it still substantially relaxes it. Moreover this is what the collider data may be hinting at, with no observation of stable narrow resonances close to the TeV scale. Nevertheless,  this heavy-KK scenario remains testable. As we will discuss hereafter, 
the present and future gravitational wave interferometers can indirectly probe KK resonances heavier than the TeV scale.

\section{The radion effective potential}
\label{sec:radion}

In this section we review for completeness details on the radion effective potential and the formalism we adopt~\cite{Megias:2020vek}.

\subsection{General formalism}

We consider the 5D action of the model~\cite{Megias:2020vek}
\begin{eqnarray}
S &=& \int d^5x \sqrt{|\det g_{MN}|} \left[ -\frac{1}{2\kappa^2} R + \frac{1}{2} g^{MN}(\partial_M \phi)(\partial_N \phi) - V(\phi) \right] \nonumber \\
&&- \sum_{\alpha} \int_{B_\alpha} d^4x \sqrt{|\det \bar g_{\mu\nu}|} \Lambda_\alpha(\phi)  + S_{\rm GHY}  \,,
\end{eqnarray}
where $S_{\rm GHY}$ is the standard Gibbons-Hawking-York boundary term, $V(\phi)$ is the bulk scalar potential, $\Lambda_\alpha(\phi)$ are the UV $(\alpha=0)$ and IR $(\alpha=1)$ 4D brane potentials located at $\phi_\alpha \equiv \phi(r_\alpha)$, and $\kappa^2 = 1/(2M_5^3)$ with $M_5$ being the 5D Planck scale. The metric is defined as in Eq.~(\ref{eq:RS_metric}) where the 4D induced metric is $\bar g_{\mu\nu} = e^{-2A(y)} \eta_{\mu\nu}$, and the Minkowski metric is given by $\eta_{\mu\nu} = \textrm{diag}(1,-1,-1,-1)$. To solve the hierarchy problem, the brane dynamics must lead to values of $\phi_0$ and $\phi_1$ yielding $A(\phi_1) - A(\phi_0) \approx \log(M_{P}/\rho)$, in analogy with Eq.~(\ref{eq:hierarchyRS}).

The classical equations of motion (EoM) in the bulk can be expressed in terms of the superpotential~$W(\phi)$ as
\begin{equation}
\hspace{-1.5cm} \phi^\prime(r) = \frac{1}{2} W^\prime( \phi) \,, \qquad A^\prime(r) = \frac{\kappa^2}{6} W(\phi) \,, \qquad V(\phi) = \frac{1}{8} \left[W^\prime( \phi) \right]^2 - \frac{\kappa^2}{6} W^2(\phi) \,. \label{eq:VW}
\end{equation}
For the brane potentials we consider $\Lambda_\alpha(\phi) = \Lambda_\alpha + \frac{1}{2} \gamma_\alpha (\phi - v_\alpha)^2$, following the Goldberger-Wise mechanism~\cite{Goldberger:1999uk} to fix the values of the scalar field at the branes equal to $v_\alpha$. As in Ref.~\cite{Goldberger:1999uk}, we work in the stiff potentials limit, where $\gamma_\alpha \to \infty$, such that $\phi(r_\alpha) = v_\alpha$. We use the freedom on $A(r_0)$ to set $A(r_0)=0$. Consequently, the relationship between the 5D and 4D Planck masses is fixed via the expression
\begin{equation}
\kappa^2 M_{P}^2=2\ell\int_0^{\bar r_1}d\bar r \, e^{-2A(\bar r)}\,, 
\end{equation}
where $\ell$ is the radius of AdS,  while $\bar r \equiv r/\ell$ and $M_{P}=2.4\times 10^{18} \, \GeV$.

\subsection{The effective potential in the warped model}
\label{subsec:effective_potential}

Using the EoM, one can express the on-shell action of the model as the 4D integral $S = - \int d^4x \,U_\eff  $ where the effective potential $U_\eff$  is 
\begin{equation}
U_\eff =  \left[ e^{-4A} \left( W + \Lambda_1 \right) \right]_{r_1}   +  \left[ e^{-4A} \left( - W + \Lambda_0 \right) \right]_{r_0}  \,. \label{eq:Ueff}
\end{equation}
We now introduce a novel technique to compute  $U_\eff$. 
Firstly, notice that the EoM for $W$ with potential $V$, cf.~Eq.~(\ref{eq:VW}), is a first order differential equation, so that it admits an integration constant that we denote by $s$. If $W_0$ is a particular solution of the EoM, then it is possible to find a general solution of the equation as an expansion of the form~\cite{Papadimitriou:2007sj,Megias:2014iwa,Lizana:2019ath}
\begin{equation}
W=\sum_{n=0}^\infty s^n W_n \,, \label{expansion}
\end{equation}
where $W_n$ can be computed iteratively from $W_0$. An explicit solution for $n=1$ is given by
\begin{equation}
W_1(\phi)=\frac{1}{\ell\kappa^2} 
\exp\left(\frac{4\kappa^2}{3}\int^\phi \frac{W_0(\tilde\phi)}{W'_0(\tilde\phi)} d\tilde\phi \right) \,.
\end{equation}
We can similarly expand the scalar field $\phi$ and the warp factor $A$ as
\begin{equation}
\phi(r) =\phi_0(r) +  s \, \phi_1(r)+\mathcal O(s^2) \,, \qquad A(r) = A_0(r)+ s \, A_1(r)+\mathcal O(s^2) \,.
\end{equation}

The integration constant $s$ is fixed by the boundary condition~$\phi(r_1) = v_1$ leading to 
\begin{equation}
s(r_1)=\frac{v_1-\phi_0(r_1)}{\phi_1[\phi_0(r_1)]}  \,.   \label{eq:s-fixing}
\end{equation}
Therefore, the superpotential acquires an explicit dependence on the brane distance $r_1$, i.e.
\begin{equation}
W(v_\alpha)=W_0(v_\alpha)+s(r_1) W_1(v_\alpha) + \cdots   \,, \qquad \textrm{with} \qquad \alpha = 0, 1\,,
\end{equation}
which in turn creates a non-trivial dependence on $r_1$ of the effective potential of Eq.~(\ref{eq:Ueff}).
Finally, the effective potential normalized to its value at $r_1 \to \infty$ is given by 
\begin{equation}
\hspace{-2.5cm} U_\eff(r_1) = [\Lambda_1+W_0(v_1)]e^{-4 A_0(r_1)}[1-4A_1(r_1)s(r_1)] + s(r_1)\left[e^{-4 A_0(r_1)}W_1(v_1)-W_1(v_0)  \right] \label{eq:Ueff2}
\end{equation}
where $\Lambda_1$ is the tension at the IR brane. In this expression we have retained terms up to first order in $s(r_1)$, an expansion which has been proven \textit{a  posteriori} to be valid.

The next step is to apply this technique to the warped phenomenological model given by
\begin{equation}
W_0(\phi) = \frac{6}{\ell \kappa^2} + \frac{u}{\ell}\phi^2  \,,
\end{equation}
with $\ell \sim {\mathcal O}(M_5^{-1})$.  In this scenario, the scalar field and warp factor turn out to be
\begin{equation}
\bar\phi_0(r) =  \bar v_0 \, e^{u \bar r} \,, \qquad A_0(r) = \bar r  + \frac{\bar v_0^2}{12} \left( e^{2u\bar r} - 1 \right) \,,
\end{equation}
where we have defined the dimensionless quantities $\bar\phi\equiv\kappa\phi$ and $\bar v_a \equiv \kappa v_a$. A solution of the EoM is also $W(\phi) = W_0(\phi) + s W_1(\phi) + \cdots$, with
\begin{equation} 
W_1(\phi) = \frac{1}{\ell \kappa^2} \left( \frac{\phi}{v_0} \right)^{4/u} e^{\kappa^2 (\phi^2 - v_0^2)/3}  \,.
\end{equation}
This leads to ${\mathcal O}(s)$ corrections in $\phi(r)$ and $A(r)$ via the EoM of Eq.~(\ref{eq:VW}).

\begin{figure*}[t]
\centering
 \begin{tabular}{c@{\hspace{3.5em}}c}
 \includegraphics[width=0.43\textwidth]{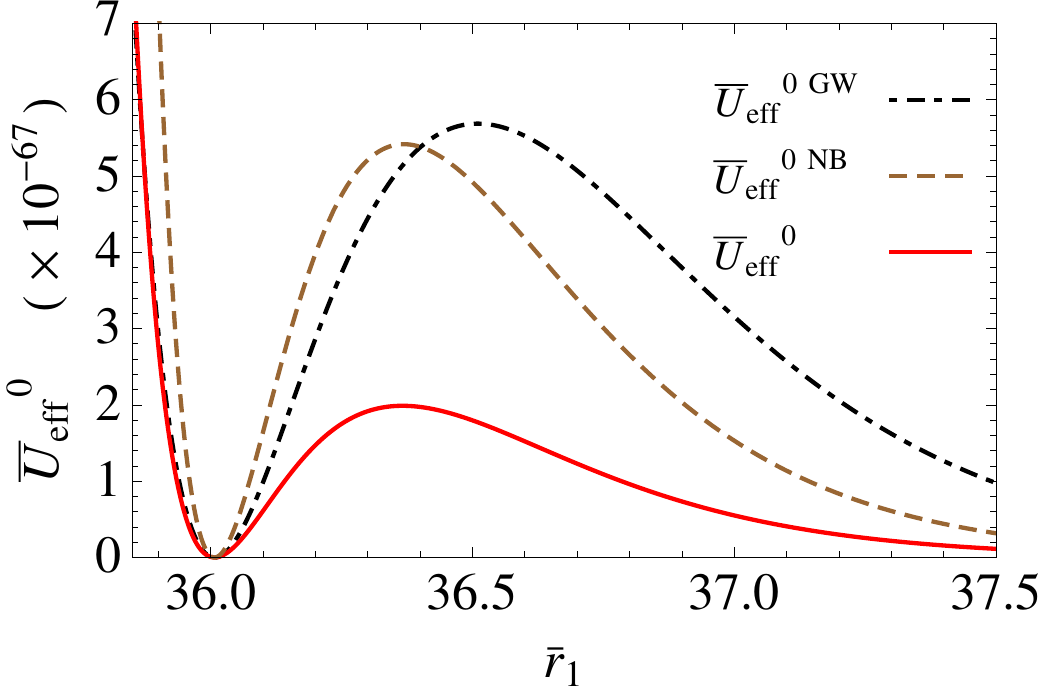} &
\includegraphics[width=0.43\textwidth]{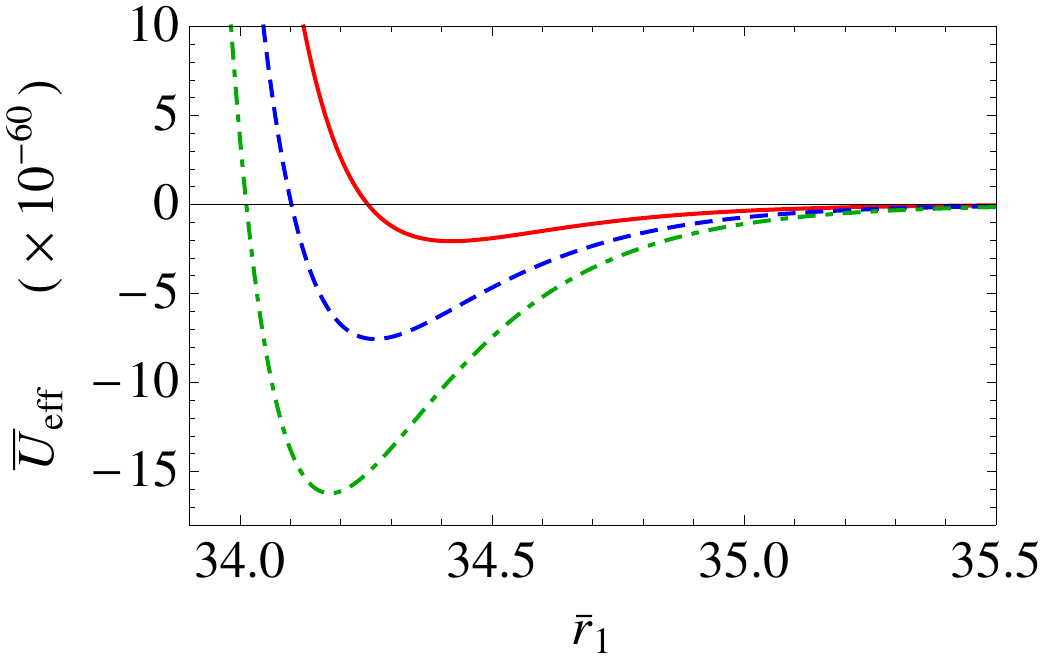} 
\end{tabular}
 \caption{\small \it Left panel: The potential $\bar U_\eff^0$ in
   Eq.~(\ref{eq:Ueff_tuned}) (solid line), the same potential when
   neglecting the backreaction $\bar U_\eff^{0 \rm NB}$ (dashed line),
   and the Goldberger-Wise potential $\bar U_\eff^{0 \rm GW}$ of
   Ref.~\cite{Goldberger:1999uk} (dashed-dotted line) as functions
   of~$\bar r_1$. Right panel: The potential $\bar U_\eff$ in
   Eq.~(\ref{eq:Ueff_detuned}) for $\lambda_1=-1$ (red solid line), $\lambda_1=-2$
   (blue dashed line) and $\lambda_1=-3$ (green dashed-dotted line). In both
   panels we assume $\bar v_0=1,\, \bar v_1=2,\, u=0.0192$\,.}
\label{fig:Ueff}
\end{figure*}

Let us first consider the simple case in which we tune to zero the first term of Eq.~(\ref{eq:Ueff2}), namely we choose $\Lambda_1 = -W_0(v_1)$. Then, by defining the dimensionless effective potential  $\bar U_\eff(r_1) \equiv \ell \kappa^2 U_\eff(r_1)$, we find
\begin{equation}
\bar U_\eff^0(\bar r_1) = 2 u^2 \bar v_1^2 (\bar r_1^0-\bar r_1) \left[e^{4A_0(\bar r_1^0)-4A_0(\bar r_1)}-1\right]\, e^{-4 A_0(\bar r_1)} \,, \label{eq:Ueff_tuned}
\end{equation}
with $v_1 \equiv v_0 \, e^{u \bar r_1^0}$. This effective potential is positive definite, and it vanishes when $u=0$, i.e. in the absence of backreaction.  In the left panel of Fig.~\ref{fig:Ueff} we display the tuned effective potential of Eq.~(\ref{eq:Ueff_tuned}) as well as, for the sake of comparison, the Goldberger-Wise potential of Ref.~\cite{Goldberger:1999uk}, denoted by
$\bar U_\eff^{0\; \rm GW}$. The agreement turns out to be much better when considering $\bar U_\eff^{0}$ with an artificially neglected backreaction on the metric, $\bar U_\eff^{0 \rm NB}$, i.e. using $A_0(r) \simeq \bar r$. Notice that $\bar U_\eff^{0\; \rm GW}$ and $\bar U_\eff^{0\; \rm NB}$ have been computed using completely different methods.

The tuned potential has degenerate minima at $\bar r_1 = \bar r_1^0$ and $\bar r_1 = + \infty$, a fact that prevents a confinement/deconfinement phase transition in an expanding universe. However, this transition is allowed in the detuned potential, which is obtained when using the condition  $\Lambda_1 + W_0(v_1) \equiv \frac{6}{\ell \kappa^2} \lambda_1 \ne 0$, where $\lambda_1$ is a dimensionless parameter. The detuned effective potential is given by
\begin{equation}
\bar U_\eff(\bar r_1) \simeq\bar U^0_\eff(\bar r_1)+6\lambda_1 e^{-4 A_0(\bar r_1)}  \,, \label{eq:Ueff_detuned}
\end{equation}
which has a minimum located at
\begin{equation}
\bar r_1^{\m} = \bar r_1^0 -\frac{1}{4}\mathcal W\left[-\frac{6\lambda_1}{u^2 \bar v_1^2}\right]  \,,
\end{equation}
where $\mathcal W(z)$ is the Lambert function. In the right panel of Fig.~\ref{fig:Ueff} we plot the effective potential in Eq.~(\ref{eq:Ueff_detuned}) for various values of $\lambda_1$. Notice that for $\lambda_1 < 0$ the global minimum is at $\bar r_1^{\m} < \bar r_1^0$.

\subsection{The radion field}
\label{subsec:radion}

The radion field is obtained by considering a scalar perturbation of the metric, i.e.~$ds^2 = e^{-2A(r) - 2F(x,y)} \eta_{\mu\nu} dx^\mu dx^\nu - [1 + 2F(x,r)]^2 dr^2$ with $F(x,r) = F(r) \mathcal R(x)$. The Lagrangian for the canonically normalized radion field $\chi(r)$ writes as
\begin{equation}
\mathcal L_\rad = \frac{6\ell^3}{\kappa^2}\int d^4x \sqrt{| \det \bar g_{\mu\nu} |} \bigg(\frac{1}{2} (\partial_\mu \chi)^2-  V_{\rm rad}(\chi) \bigg) \,,
\end{equation}
where $V_\rad(\chi)  \equiv U_\eff[\bar r_1(\chi)]  \simeq \frac{1}{2}m_\rad^2\,\chi^2$. To leading approximation in the parameter $u$, one finds
\begin{equation}
\chi(r_1) \simeq \frac{1}{\ell} e^{-A_0(r_1) } \,.
\end{equation}
The radion potential has a minimum at~$\langle\chi\rangle = \rho$, with $\rho \equiv \frac{1}{\ell} e^{-A_0(\bar r_1^m)}$. The energy scale set by $\rho$ is the physically relevant parameter that  is usually considered at the $\TTeV$ scale. $\rho$ is mainly determined by $u$, so that for instance we find $u\simeq 0.0192 \, (0.0219)$ for $\rho=1\, (100)$~TeV.

\begin{figure}[t]
\centering 
\includegraphics[width=7cm]{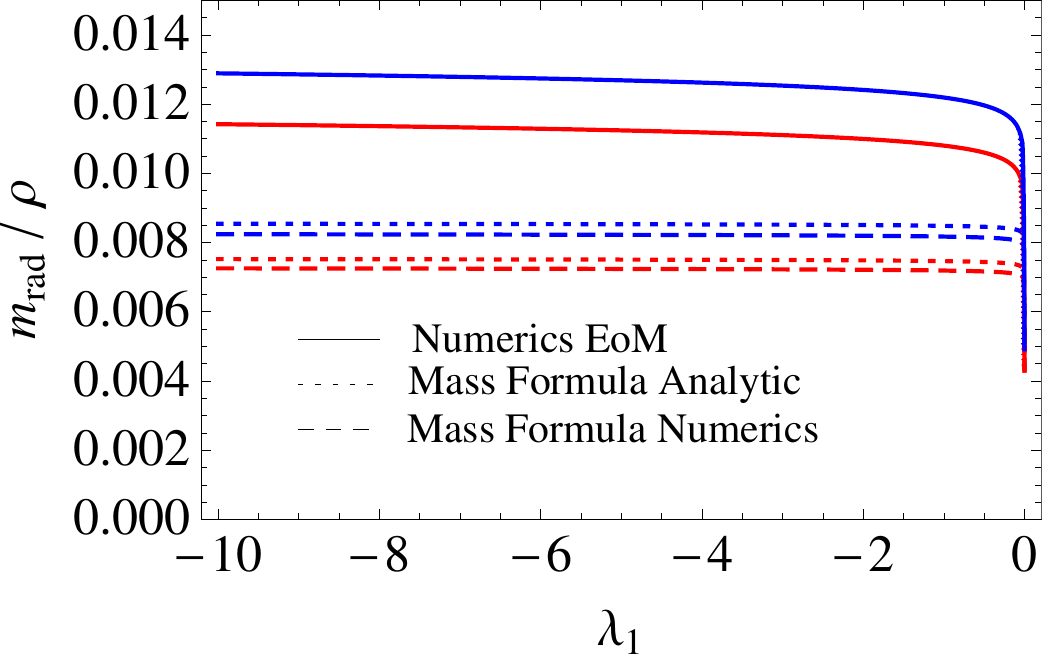}  \hspace{0.9cm} \includegraphics[width=7cm]{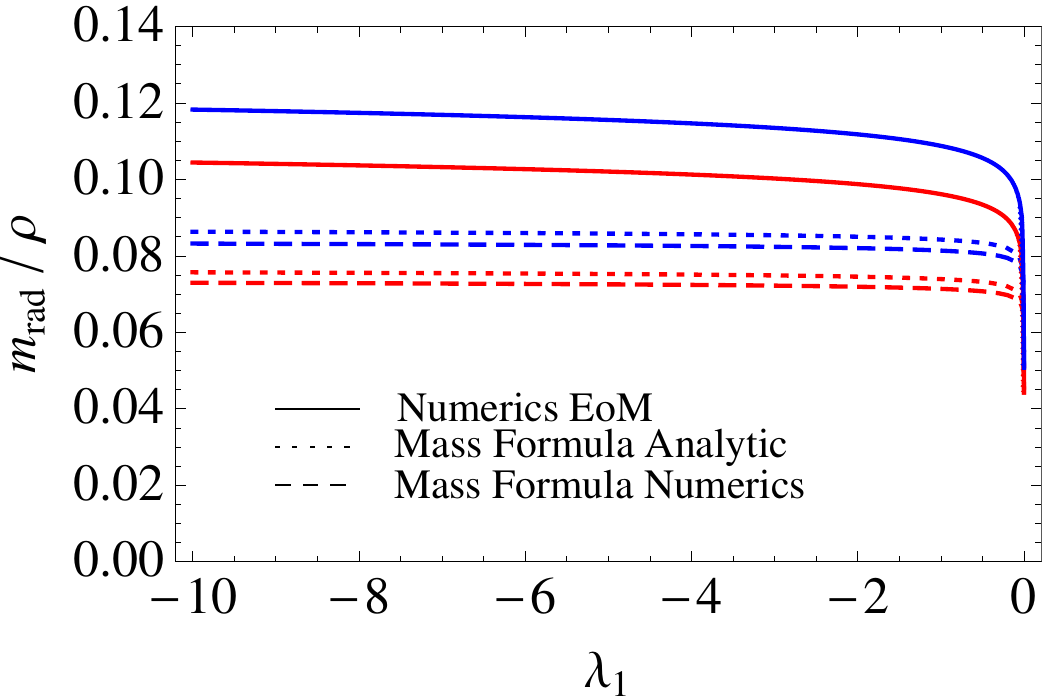}  \\  
\caption{\small \it The normalized radion mass $m_{\rm rad}/\rho$ as a function of $\lambda_1$ for $\bar v_0 = 0.1$ and $\bar v_1 = 0.2$ (left panel) and for $\bar v_0 = 1$ and  $\bar v_1 = 2$ (right panel). The red and blue lines are obtained for $\rho = 1$\,TeV and $\rho = 100$\, TeV, respectively. The solid lines are evaluated by means of the exact numerical solution of the radion EoM, Eqs.~(\ref{eq:EoMF1})-(\ref{eq:EoMF2}). The results by using the mass formulas have been evaluated in two ways: i) fully numerically as it is written in Eqs.~(\ref{eq:mrad})-(\ref{eq:Pirad0}) (dashed lines), and ii) using the analytical approximation given by Eq.~(\ref{eq:masaradionapp}) (dotted lines).}
\label{fig:radionmass}
\end{figure}

The radion mass can be computed by solving the EoM of the scalar perturbation. Using the background EoM one can recast the bulk EoM and boundary conditions for the excitation $F$ as~\cite{Megias:2015ory}
\begin{eqnarray}
&&\partial_r \left( e^{2A} A^{\prime\prime}(r)^{-1} \partial_r \left[ e^{-2A} F \right] \right) + \left( m_{\rm rad}^2 e^{2A} A^{\prime\prime}(r)^{-1} - 2 \right) F = 0 \,,  \label{eq:EoMF1} \\
&&\hspace{3.8cm} \left[ m_{\rm rad}^2 F + U_\alpha^{\prime\prime}[\phi(r)] \partial_r [e^{-2A} F] \right] \Big|_{r_\alpha} = 0 \,, \label{eq:EoMF2}
\end{eqnarray}
where the localized effective potentials are defined by $U_\alpha(\phi) = \Lambda_\alpha(\phi) - (-1)^\alpha W(\phi)$. The solution of the EoM can be computed, either numerically or by using some analytical mass formulas obtained in the light radion approximation (see e.g.~Ref.~\cite{Megias:2018sxv}). In the stiff limit of the  brane potentials, for which the radion mass is maximized, the mass formula reads
\begin{equation}
\frac{m_{\rm rad}^2}{\rho^2} =  \frac{1}{\Pi_{\rm rad}} \,,
\label{eq:mrad}
\end{equation}
with
\begin{eqnarray}
\Pi_{\rm rad} &=& \frac{1}{\ell^2}\int_0^{r_1^{\textrm{\tiny m}}}dr e^{4(r-r_1^{\textrm{\tiny m}})/\ell}e^{4[\Delta A(r)-\Delta A(r_1^{\textrm{\tiny m}})]}\left(\frac{W[\phi(r)]}{W'[\phi(r)]} \right)^2\nonumber\\
&&\qquad \times \left[   \frac{2}{W[\phi(r_1^{\textrm{\tiny m}})]}+\int_r^{r_1^{\textrm{\tiny m}}}d\bar r e^{-2[A(\bar r)-A(r_1^{\textrm{\tiny m}})]} \left(\frac{W'[\phi(\bar r)]}{W[\phi(\bar r)]} \right)^2
\right] \,, \label{eq:Pirad0}
\end{eqnarray}
where $\Delta A(r)\equiv A(r)-\bar r$. The integral in Eq.~(\ref{eq:Pirad0}) is dominated by the region $r\simeq r_1^\m$. Its analytical approximation in the limit of $u\ll 1$ is given by~\footnote{The result follows from the approximation 
$$
I_G \equiv \int_0^{r_1^{\textrm{\tiny m}}} dr \, e^{4(r-r_1^m)/\ell} G(r) \simeq \frac{\ell}{4} G(\bar r_1^{\textrm{\tiny m}} - 1/4) \,, \label{eq:Theorem}
$$
valid in the regime $\bar r_1^{\textrm{\tiny m}} \gg 1$. To check this out, one performs a Taylor expansion of $G(r)$ around $r = r_1^{\textrm{\tiny m}}$. Then, the integrand in the left-hand side can be easily integrated, and compared order by order with the Taylor expansion of $\frac{\ell}{4}G(\bar r_1^{\textrm{\tiny m}} -x)$ around $x = 1/4$. When neglecting contributions of order $\mathcal{O}(e^{-4 \bar r_1^{\textrm{\tiny m}}})$, both sides of the equation coincide at leading and next-to-leading order in the Taylor expansion, and disagree at the next order, since
$
I_G = \ell G (\bar r_1^{\textrm \tiny m})/4 - \ell^2 G^\prime(\bar r_1^\textrm{\tiny m})/16  + \ell^3/64 \;\mathcal{O}(G^{\prime\prime}(\bar r_1^{\textrm{\tiny m}})) \,.
$
}
\begin{equation}
\Pi_{\rm rad}\simeq \frac{1}{4\ell} \frac{2}{W[\phi(r_1^m)]}\,f(\bar r_1^m-1/4)\,, \nonumber\\
\label{eq:Pirad}
\end{equation}
and therefore
\begin{equation}
\frac{m_{\rm rad}}{\rho} \simeq \left. e^{2[\Delta A(\bar r_1^m)- \Delta A(r)]}
\sqrt{2\ell\,W[\phi(r_1^m)]}
\,
\frac{W'[\phi(r)]}{W[\phi(r)]}\right|_{\bar r=\bar r_1^m-1/4} \,.
\label{eq:masaradionapp}
\end{equation}
The results using the numerical calculation and this mass formula are displayed in Fig.~\ref{fig:radionmass}. Both results are in reasonable agreement: for $\bar v_0 = 1$, $\bar v_1 = 2$ and $\rho = 1 \textrm{ and } 100 \, \TTeV$, the radion mass is $m_{\rm rad}/\rho \simeq 0.10$, and it scales linearly with the values of $\bar v_{0,1}$, while it is almost independent of $\lambda_1$. 

Using similar approaches, we can compute the mass of the first KK resonance for gauge bosons and gravitons, and we obtain $m_{\rm KK}^{\rm gauge}/\rho \simeq 2.46$ and $m_{\rm KK}^{\rm grav}/\rho \simeq 3.88$, values almost independent of $\bar v_{0,1}$, $\lambda_1$ and $\rho$. Then, we find a hierarchy of scales
$m_\rad  \ll  m_\KK^{\rm gauge} \lesssim  m_\KK^{\rm grav}  $,
which guarantees that the theory is perturbative.

\section{The phase transition}
\label{sec:Effective_Potential_FiniteT}

In the AdS/CFT correspondence a black hole (BH) solution describes the high temperature phase of the theory. In this phase  the dilaton is in the symmetric phase $\langle \chi\rangle = 0$. The BH metric is
\begin{equation}
ds_\BH^2 = - h(r)^{-1} dr^2 + e^{-2A(r)} (h(r) dt^2 -  d\vec{x}^{\,2} ) \,, \label{eq:metricBH}  
\end{equation}
where the blackening factor $h(r)$ vanishes at the event horizon, i.e.~$h(r_h) = 0$, and satisfies the boundary condition $h(0) = 1$. The solution of the EoM in the $u \ll 1$ limit leads to
\begin{equation}
h(r)\simeq 1-e^{4[A_0(r)-A_0(r_h)]}\,.
\end{equation}
Then, the Hawking temperature and the minimum of the free energy in the BH solution read as
\begin{equation}
T_h = \frac{1}{4\pi} e^{-A(r_h)} \big| h^\prime(r) \big|_{r=r_h}  \simeq \frac{1}{\ell \pi} e^{-A_0(r_h)} \,, \qquad F_{\min}^\BH(T)\simeq -\frac{\pi^4\ell^3}{\kappa^2}T^4\,.   \label{eq:Ths}
\end{equation}

Below a given temperature, a competing phase with  $\langle \chi \rangle \ne 0$ arises. This is the confined phase. 
The free energies of the BH deconfined phase and the  
confined phase  are
\begin{equation}
F_{d}(T) = E_0 + F_{\min}^\BH - \frac{\pi^2}{90}g_d^\eff T^4 \,, \qquad F_{c}(T)=-\frac{\pi^2}{90}g_c^\eff T^4 \,,
\end{equation}
where $E_0 = V_\rad(0) - V_\rad(\rho) > 0$ is the $T=0$ potential gap between the two phases. For concreteness we assume~$g_c^\eff \simeq g_d^\eff \simeq 100$. The phase transition can start when $F_d \le F_c$, so that we can define the critical temperature as $F_d(T_c) = F_c(T_c)$. Since the two phases are separated by a potential barrier, the phase transition proceeds through bubble nucleation of the confined phase in the deconfined sea. It happens when the barrier between the false BH minimum and the true vacuum is overcome with sufficiently high probability. While at high $T$ this process is driven by thermal fluctuations and the corresponding Euclidean action is $O(3)$ symmetric, at low $T$ the transition happens via quantum fluctuations with an $O(4)$ symmetric action.

The Euclidean action with symmetry $O(n)$ is given by
\begin{equation}
S_n= \Omega_n \int d\sigma \sigma^{n-1} \frac{6\ell^3}{\kappa^2} \left(\frac{1}{2} \left( \frac{\partial \chi}{\partial \sigma} \right)^2 + V(\chi,T)\right) \,,
\label{eq:Euclidean_Action}
\end{equation}
with 
\begin{equation}
V(\chi,T) \equiv\frac{\kappa^2}{6\ell^3} \left( V_{\rad}(\chi) +   \left|F_{\min}^{\BH}(T)\right|   \right) \,.
\end{equation}
The quantity $\Omega_n = n \pi^{n/2}/\Gamma(1 + n/2)$ is the surface of the unit $n$-sphere, and  we define $\sigma = \sqrt{\vec{x}^2}$ and $\sigma = \sqrt{\vec{x}^2 + \tau^2}$ for $n=3$ and $n=4$, respectively. Here $\tau$ is the Euclidean time. The corresponding EoM is known as the ``bounce equation'', and it is written as~\cite{Coleman:1977py,Linde:1980tt}
\begin{equation}
\frac{\partial^2 \chi}{\partial\sigma^2}  + \frac{(n-1)}{\sigma} \frac{\partial \chi}{\partial\sigma} - \frac{\partial V}{\partial\chi} = 0  \,,  \label{eq:bouncesol3} 
\end{equation}
with boundary  conditions~$\chi(0)=\chi_0$ and $d\chi/d\sigma\big|_{\sigma = 0} = 0$. The corresponding temperature is obtained by equating the kinetic energy of the radion at the origin, $\chi = 0$, with the thermal energy, so that the barrier between the two vacua can be overcome, i.e.
\begin{equation}
\frac{3\ell^3}{\kappa^2} \left( \frac{\partial \chi}{\partial \sigma} \right)^2 \Bigg|_{\chi = 0} = \left|F_{\min}^{\BH}(T)\right| \,. 
\end{equation}

The bubble nucleation rate from the false BH minimum to the true vacuum per Hubble volume $\mathcal V$ is~$\Gamma/\mathcal V \sim  T_c^4 \left( e^{-S_3/T} + e^{-S_4} \right)$, so that it is dominated by the least action. The onset of the transition occurs at the nucleation temperature~$T_n$, with $T_n < T_c$, defined as the temperature at which  the probability for a single bubble to be nucleated within one horizon volume is $\mathcal O(1)$. This corresponds to~\cite{Konstandin:2010cd}
\begin{equation}
S_E(T_n) \lesssim 4 \log \frac{M_P/\rho}{ T_n/\rho} \; \approx  \left\{ 
\begin{array}{cc}
140 & \qquad  \textrm{for} \qquad \rho \approx 1 \, \textrm{TeV}   \\
120 & \qquad \quad  \textrm{for} \qquad \rho \approx 100 \, \textrm{TeV} 
\end{array} \,. \right. 
\label{eq:SE_bound}
\end{equation}

In the regime where bubble formation is dominated by thick wall bubbles~\cite{Randall:2006py}, one may consider the thick wall approximations~\cite{Bunk:2017fic}:
\begin{eqnarray}
&& \frac{S_3(T)}{T} \simeq \frac {\sqrt{3}N^2 }{ (\pi T/\rho) \sqrt{{\mathcal V}(T/\rho)} } \,, \qquad S_4(T)\simeq\frac{9N^2}{4{\mathcal V}(T/\rho)} \,,
\label{eq:actions_thickwall}
\end{eqnarray}
where  $N$ is the number of colors in the dual CFT theory, and $\mathcal V(T/\rho)= \pi^4(T_c^4 - T^4)/\rho^4 $.
By means of the AdS/CFT duality, $N$ is connected to the 5D squared gravitational coupling constant $(M_5 \ell)^{-3}$ via the relationship $N^{2} = (M_5 \ell)^{3}16\pi^2$, so that the gravitational theory is weakly coupled in the limit $N\gg 1$. However, from expressions (\ref{eq:actions_thickwall}) we see that in the limit $N\to\infty$ the actions diverge and there is no phase transition at all. At finite $N$, instead, the phase transition occurs and the tunneling temperature depends on the value of $N$ (see next section). Although the formulas of Eq.~(\ref{eq:actions_thickwall}) are quite intuitive, we find that the thick wall approximation often mismatches the fully numerical result.
The nucleation temperatures calculated with one or the other method can differ by almost an order of magnitude. We do not investigate further the reason, as computing $S_3/T$ and $S_4$ by solving numerically the bounce equation~(\ref{eq:bouncesol3}) is not very demanding. We mostly comment on this aspect to rise awareness on the issue.

\begin{figure*}[t]
\centering
 \begin{tabular}{c@{\hspace{3.5em}}c}
 \includegraphics[width=0.43\textwidth]{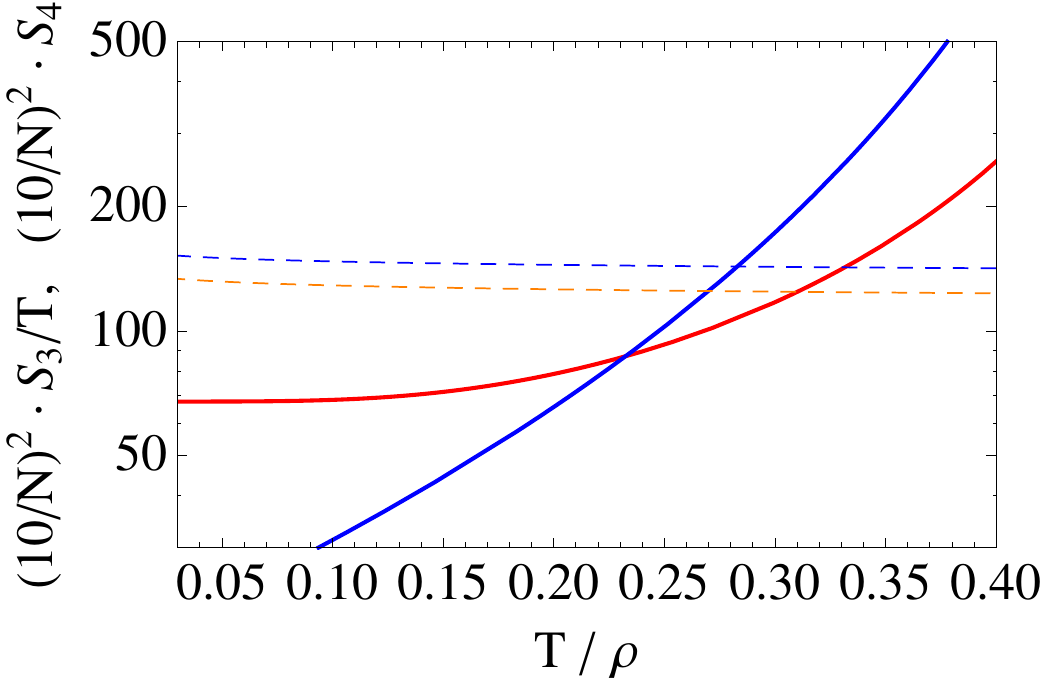} &
\includegraphics[width=0.43\textwidth]{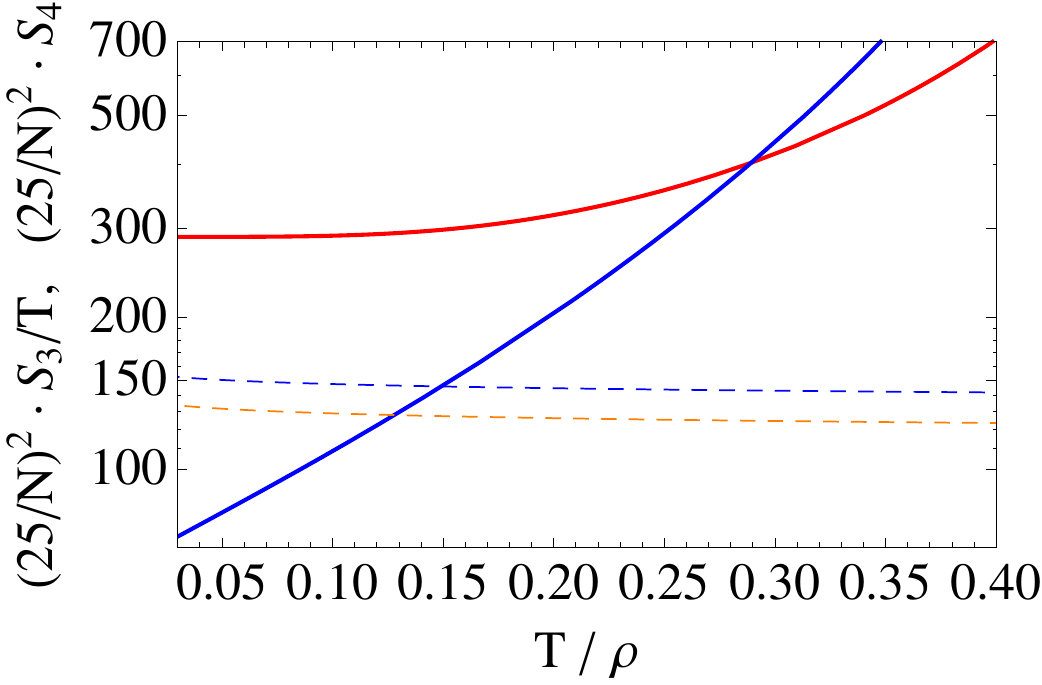} 
\end{tabular}
 \caption{\small \it The Euclidean actions $S_3 / T$ (red) and $S_4$ (blue) as functions of $T$ computed by solving numerically the bounce equation~(\ref{eq:bouncesol3}). We display the cases of $\lambda_1 = -3.0$ (left panel) and $\lambda_1 = -5.0$ (right panel). Horizontal blue (orange) lines correspond to the bounds for $\rho=1$~TeV (100 TeV), cf. Eq.~(\ref{eq:SE_bound}).}
\label{fig:SE}
\end{figure*}

In Fig.~\ref{fig:SE} we display the Euclidean actions as functions of the temperature computed numerically. In the left panel we present a scenario where the phase transition is dominated by the ${\mathcal O}(3)$  contribution, while in the left panel by the ${\mathcal O}(4)$ contribution. The curves are normalized by the number of colors in the dual theory. 

It is useful to notice that a short inflationary epoch precedes the phase transition, and some reheating is realized after it.  Inflation starts at the temperature $T_i$ such that $E_0$ dominates the value of energy density in the deconfined phase, $\rho_d$, over the thermal corrections, where
\begin{equation}
\rho_{d}(T) = E_0 + \frac{3\pi^4\ell^3}{\kappa^2}T^4 + \frac{\pi^2}{30}g_d^\eff T^4 \,. \label{eq:E0}
\end{equation}
\begin{figure*}[b]
\centering
 \begin{tabular}{c@{\hspace{3.5em}}c}
 \includegraphics[width=0.43\textwidth]{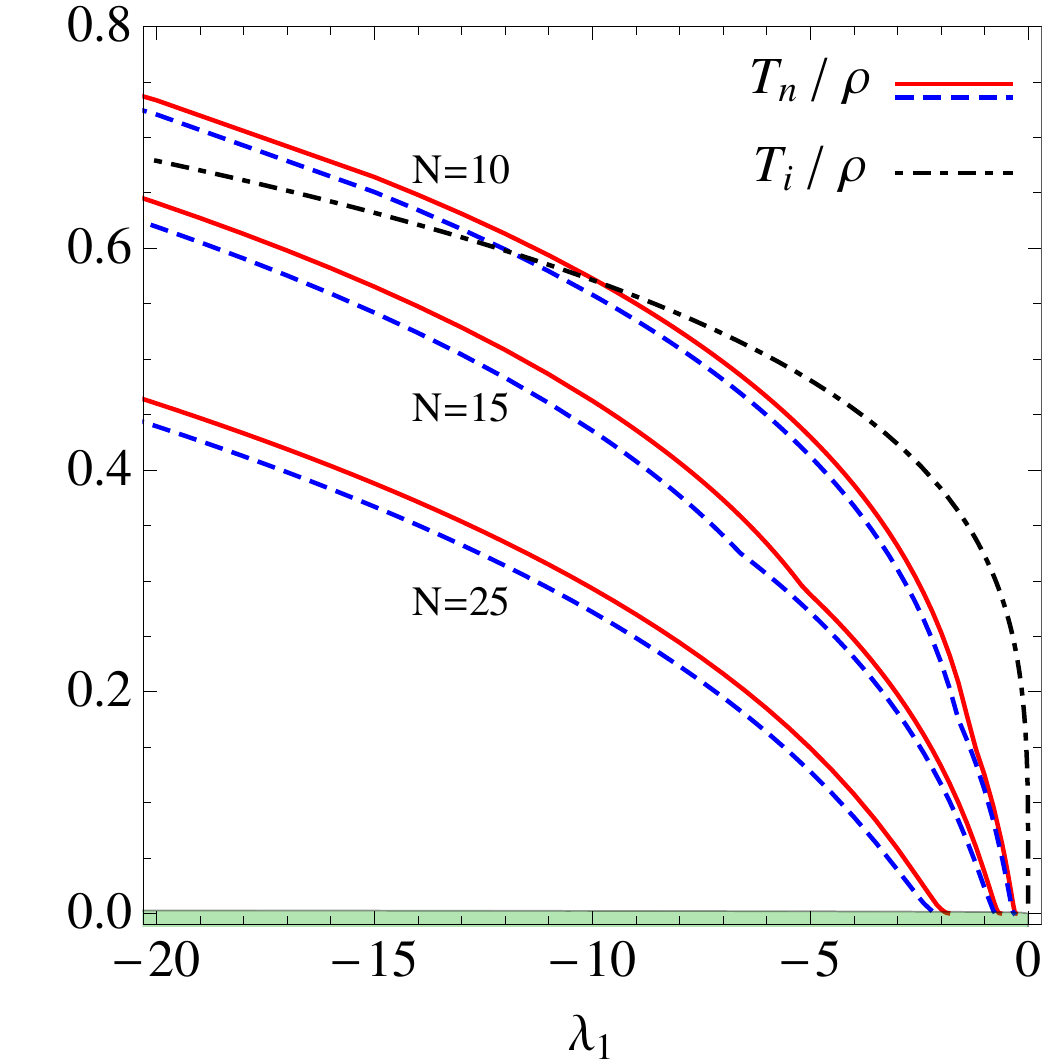} &
\includegraphics[width=0.43\textwidth]{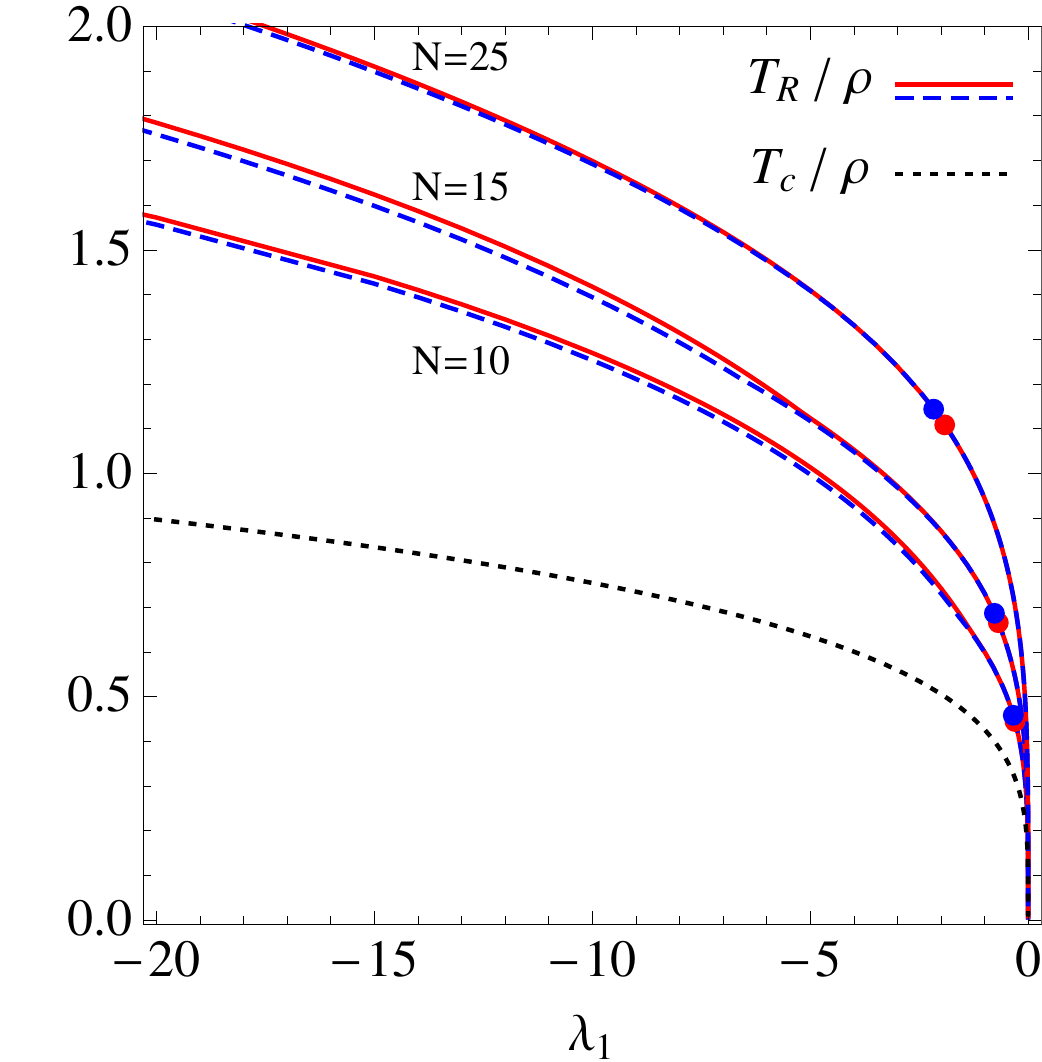} 
\end{tabular}
 \caption{\small \it The normalized temperatures $T_n/\rho$ (left panel, solid and dashed curves),
   $T_i/\rho$ (left panel, dashed-dotted black curve), $T_R/\rho$
   (right panel, solid and dashed curves) and $T_c/\rho$ (right panel,
   dotted black curve) as a function of $\lambda_1$ for different
   values of $N$ and $\rho$. Solid-red and dashed-blue lines
   correspond to the scenarios with $\rho=1$\,TeV and $\rho=100$\,TeV, respectively.}
\label{fig:Tn_TR}
\end{figure*}
After the phase transition, $\rho_d$ is converted into radiation density in the confined phase. The temperature then goes up to the reheat temperature $T_R$, which is obtained from the requirement $\rho_c(T_R)=\pi^2 g_c^\eff T_R^4/30\simeq \rho_d(T_n)$.  The gravitational radiation is a subleading contribution. Fig.~\ref{fig:Tn_TR} shows the behavior of $T_n$ and $T_i$ (left panel) as well as $T_R$ and $T_c$ (right panel) as functions of $\lambda_1$ for different values of $N$ and~$\rho$. As expected, the condition $T_R > T_c$ is always satisfied. The big bang nucleosynthesis (BBN) bound excludes the shadowed (green) region on the bottom of the left panel as $T_n/\rho \gtrsim 3 \times 10^{-4} \sqrt{N}$~\cite{Megias:2020vek}. The circles in the right panel correspond to lower bounds of parameter configurations on the border of the BBN bound.

\section{The stochastic gravitational waves background}
\label{sec:Gravitational_waves}

A cosmological first order phase transition generates a SGWB with a power spectrum $\Omega_\GW(f)$ depending on the dynamics of the bubbles in the plasma. Here we consider two different approaches to estimate $\Omega_\GW(f)$: the envelope approximation, $\Omega^\env_\GW(f)$, in which plasma effects are assumed to be negligible~\cite{Caprini:2015zlo, Huber:2008hg}, and the sound waves numerical result, $\Omega^\sw_\GW(f)$, in which the coherent motion of the plasma is taken into account~\cite{Caprini:2015zlo,Hindmarsh:2017gnf, Caprini:2019egz}. The power spectra in these two regimes can be approximated as
\begin{equation}
\Omega^{\env}_\GW(f)\simeq 
\frac{3.8\, x^{2.8}}{1+2.8\,x^{3.8}}\,\overline\Omega_{\rm GW}^{\,\env}  \,, \qquad \Omega^{\sw}_\GW(f)\simeq x^{3} \left( \frac{7}{4+3\,x^2} \right)^{7/2} \,\overline\Omega_{\rm GW}^{\,\rm sw} \,, 
\end{equation}
where $f_p$ is the peak frequency,  $\overline\Omega_{\rm GW}^{\,\env (\sw)}$ is the corresponding amplitude at the peak, and $x = f/f_p$. The quantities $f_p$ and $\overline\Omega_{\rm GW}^{\,\env (\sw)}$  depend on the phase transition parameters $v_w$, $\alpha$ and $\beta/H_\star$, which are respectively the bubble expansion velocity,  the normalized gap between the energies in the two phases, and the inverse time duration of the phase transition. The explicit dependence of $\overline\Omega_{\rm GW}^{\,\sw}$ and $\overline\Omega_{\rm GW}^{\,\env}$ are taken from Refs.~\cite{Caprini:2015zlo, Caprini:2019egz}, although in the former we have introduced an additional suppression factor accounting for the sound wave finite lifetime effect~\cite{Guo:2020grp,Hindmarsh:2020hop}. Its order of magnitude is inferred from Fig.~17 of Ref.~\cite{Ares:2020lbt} although, in the lack of simulations at $\alpha \gg 1$, it should be considered only as a rough estimator of the possible uncertainties affecting the SGWB prediction.

For the bubble expansion velocity, there is no unquestionable way to calculate it in the case of the radion transition, hence  we treat it as an external input and set it at $0.8\lesssim v_w\lesssim 1$ (in the scale of our plots, the variation of $v_w$ within this interval is not visible).  The energy gap and the inverse time duration  can instead be approximated as~\cite{Caprini:2015zlo, Caprini:2019egz}
\begin{equation}
  \alpha \simeq  \frac{|F_{d}(T_n) - F_{c}(T_n)|}{\rho_{d}^\ast(T_n)}   \; \qquad, \qquad
   \frac{\beta}{H_\star}\simeq 
T_n  \frac{dS_E}{dT}\bigg|_{T=T_n} \, ,
\end{equation}
where $ \rho_{d}^\ast(T_n) = \rho_d(T_n) - E_0 $.
The larger $\alpha$ and smaller $\beta/H_\star$, the stronger the SGWB signal. 

\begin{figure*}[t]
\centering
 \begin{tabular}{c@{\hspace{3.5em}}c}
 \includegraphics[width=0.41\textwidth]{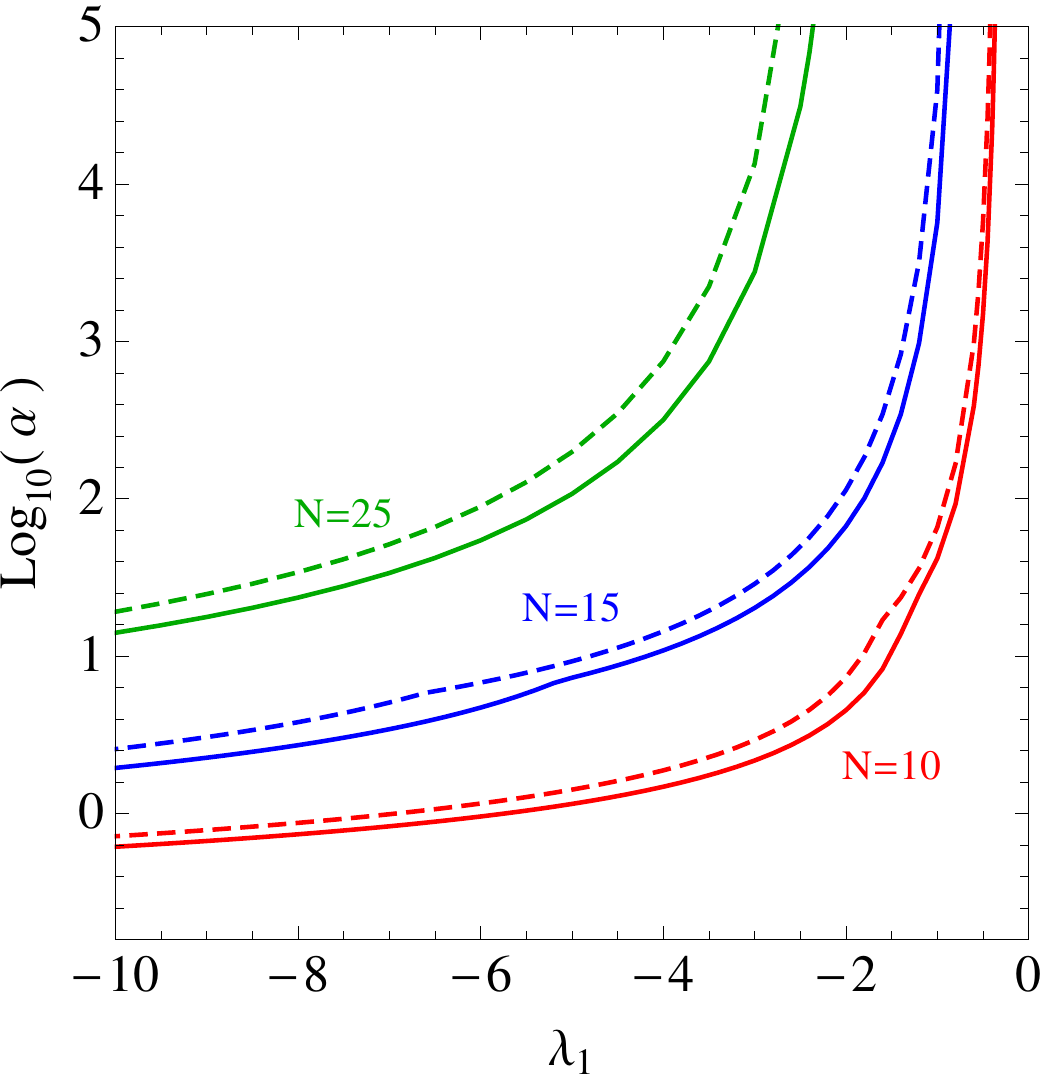} &
\includegraphics[width=0.43\textwidth]{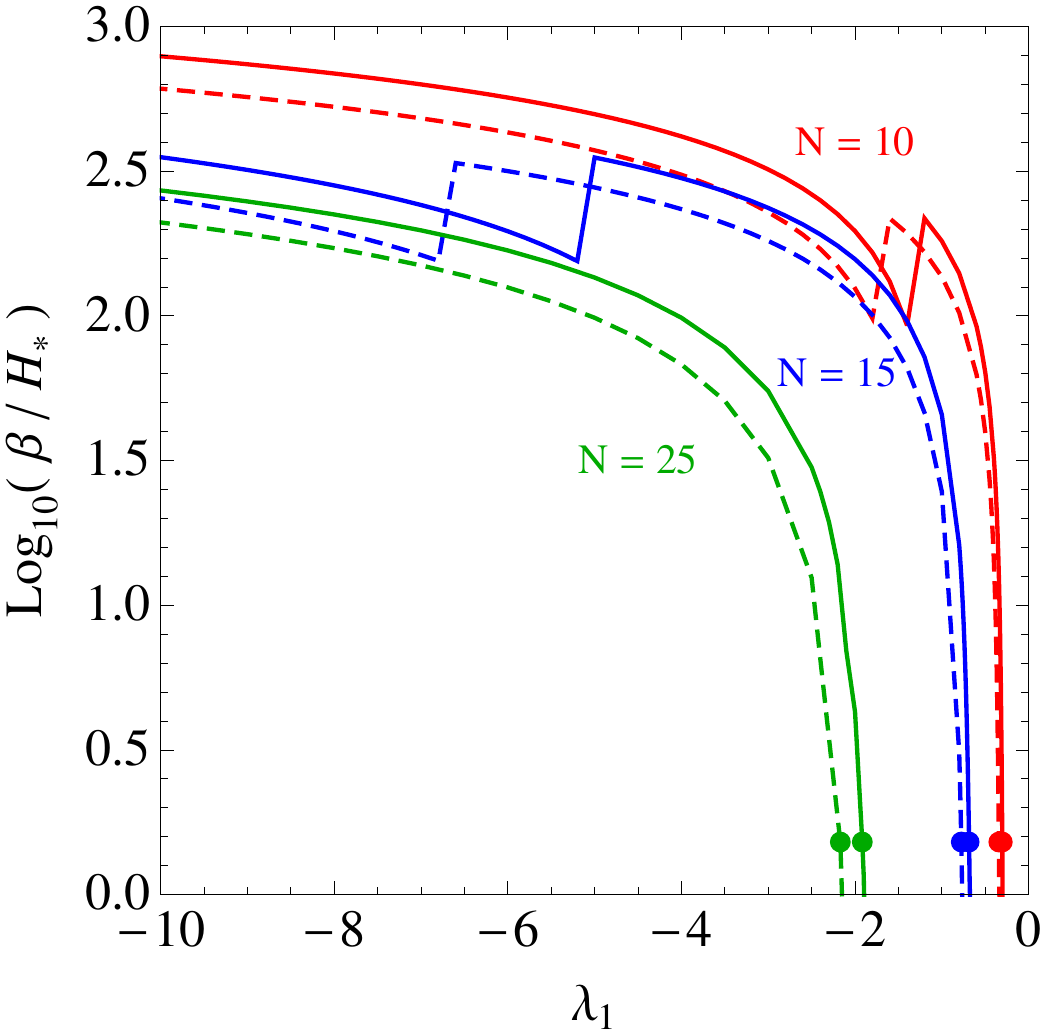} 
\end{tabular}
 \caption{\small \it The parameters $\alpha$ (left panel) and $\beta/H_\star$ (right panel)
   as functions of $\lambda_1$ for different values of $N$ and
   $\rho$. Solid and dashed lines assume $\rho=1$~TeV and $\rho=100$\,TeV, respectively.}
\label{fig:alpha_beta}
\end{figure*}
\begin{figure*}[b]
\centering
 \begin{tabular}{c@{\hspace{3.5em}}c}
 \includegraphics[width=0.43\textwidth]{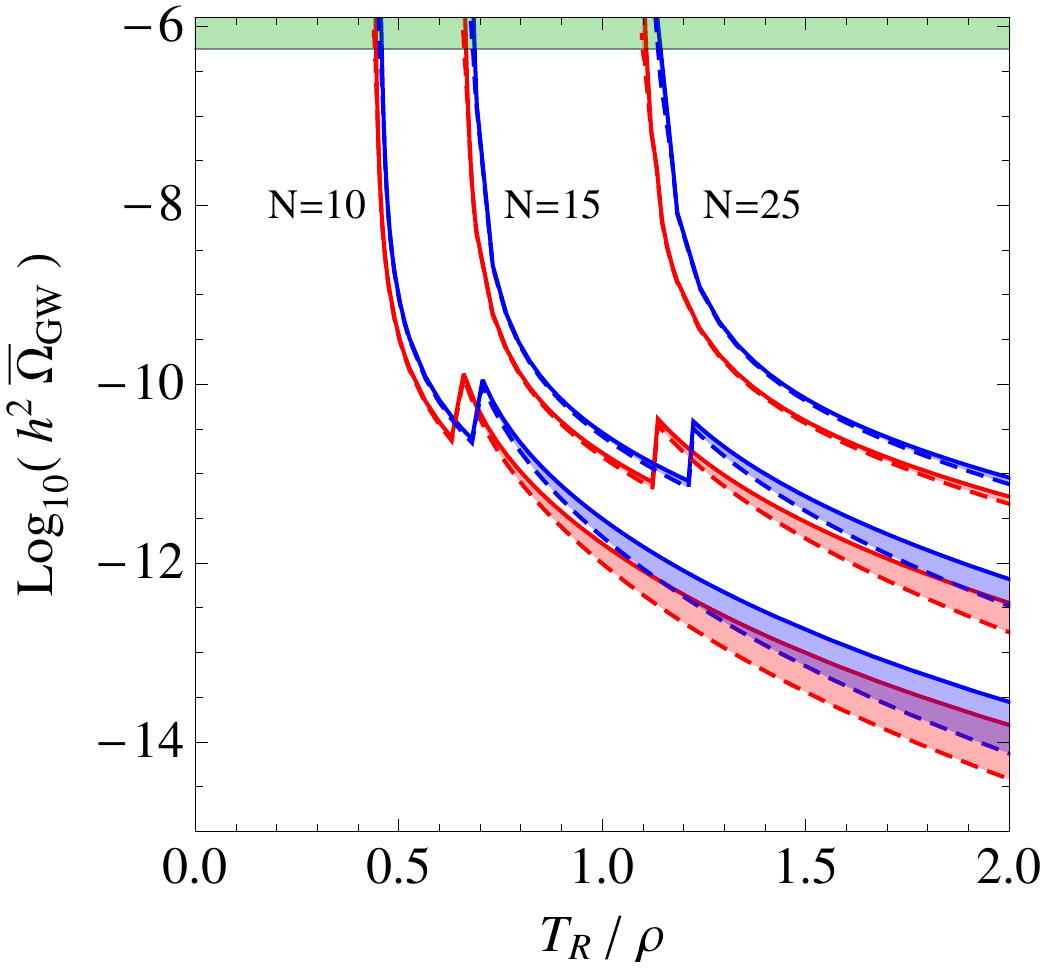} &
\includegraphics[width=0.41\textwidth]{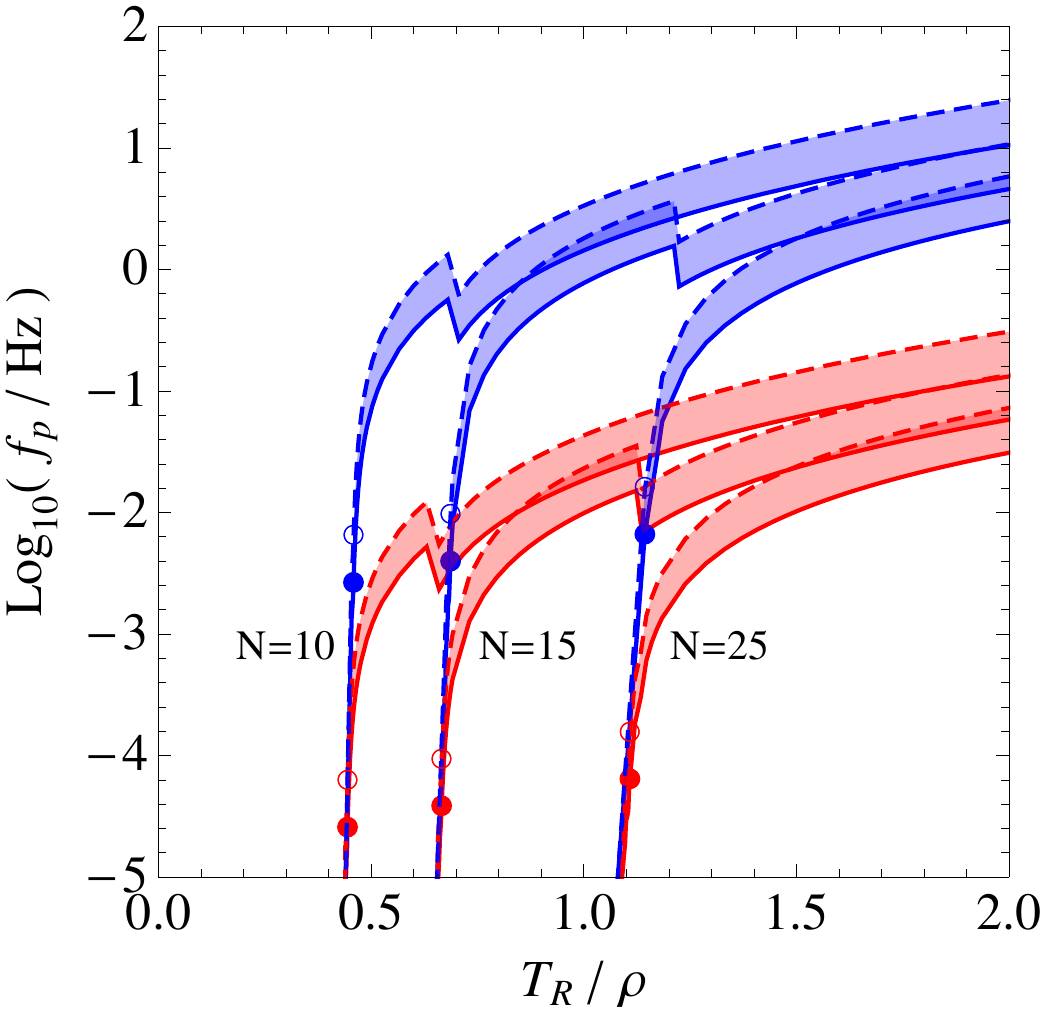} 
\end{tabular}
 \caption{\small \it The normalized maximal amplitude $h^2 \overline\Omega_{\textrm{\tiny GW}}$ (left panel) and its frequency $f_p$ (right panel) as a function of $T_R/\rho$ for different values of $N$ and $\rho$. The strips are for $\rho = 1$\,TeV (red) and $\rho = 100$\,TeV (blue). The shadowed green region in the left panel is excluded by BBN. In the right panel, the border of the BBN bound corresponds to the circles. We have considered for the wall velocity $v_\omega\simeq 0.99$.}
\label{fig:Omega_fp_TR}
\end{figure*}

\begin{figure*}[t]
\centering
 \begin{tabular}{c@{\hspace{3.5em}}c}
 \includegraphics[width=0.41\textwidth]{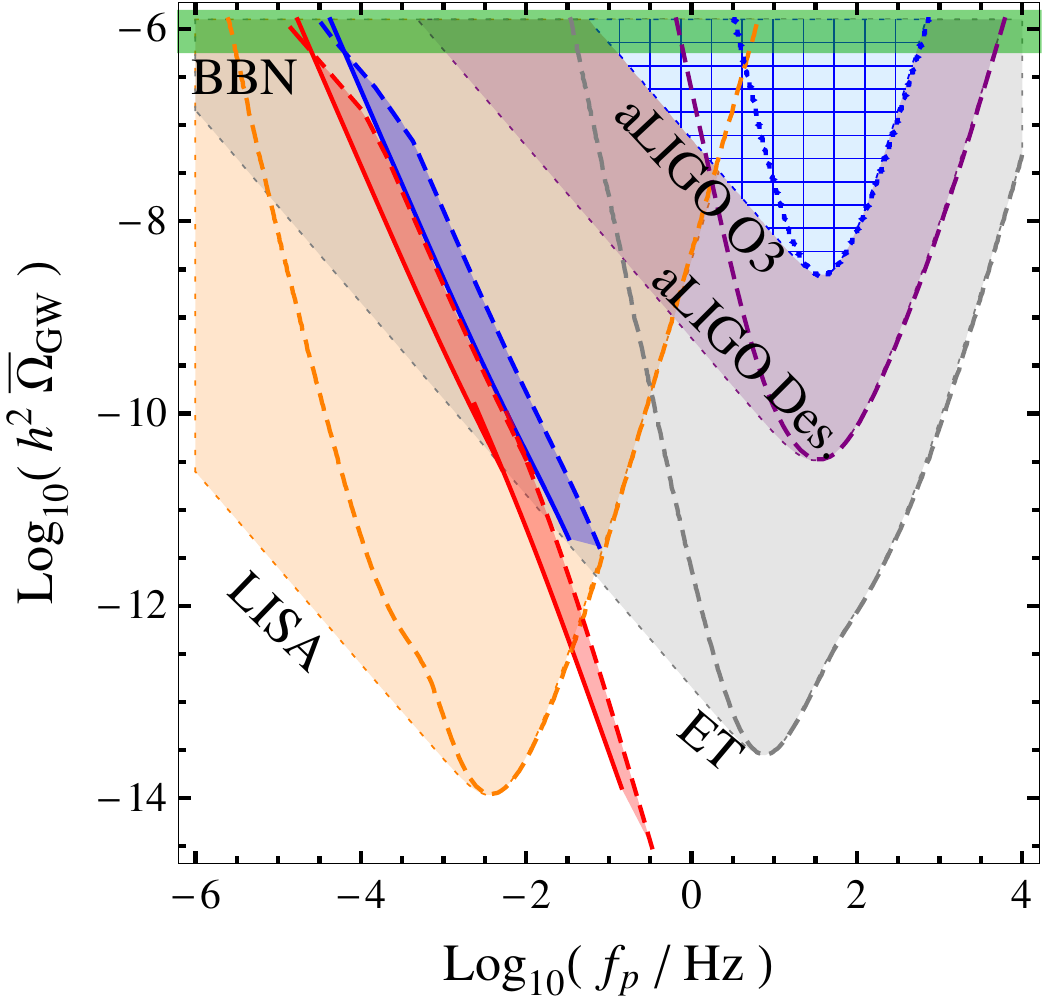} &
 \includegraphics[width=0.41\textwidth]{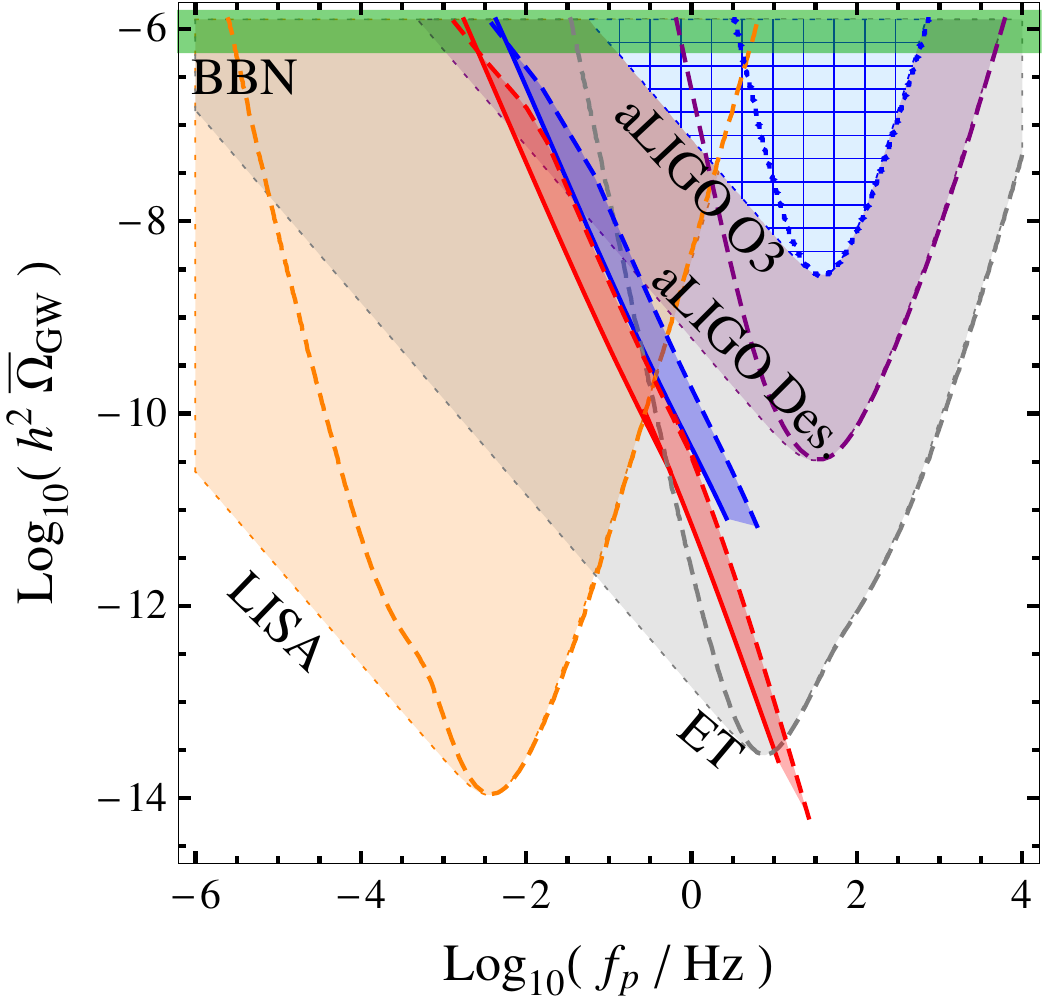} \\
 \includegraphics[width=0.41\textwidth]{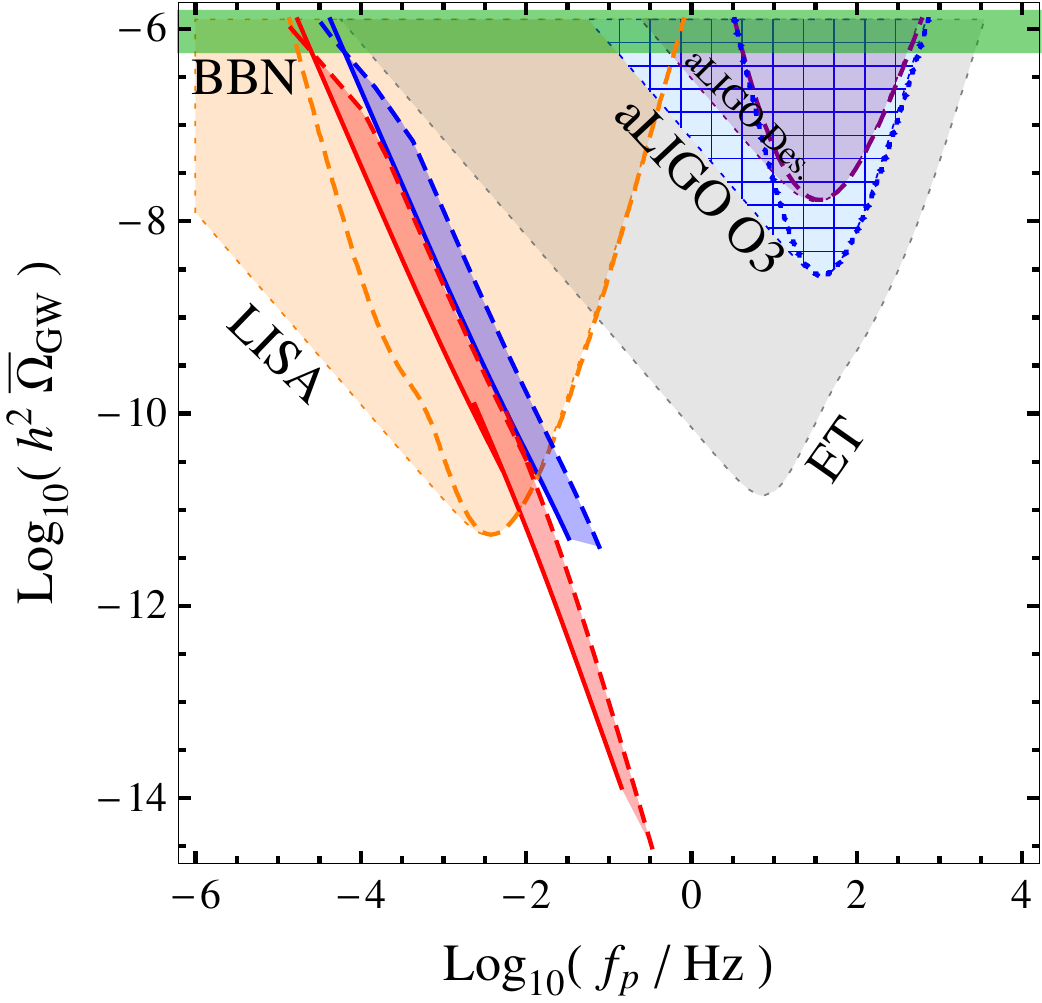} &
 \includegraphics[width=0.41\textwidth]{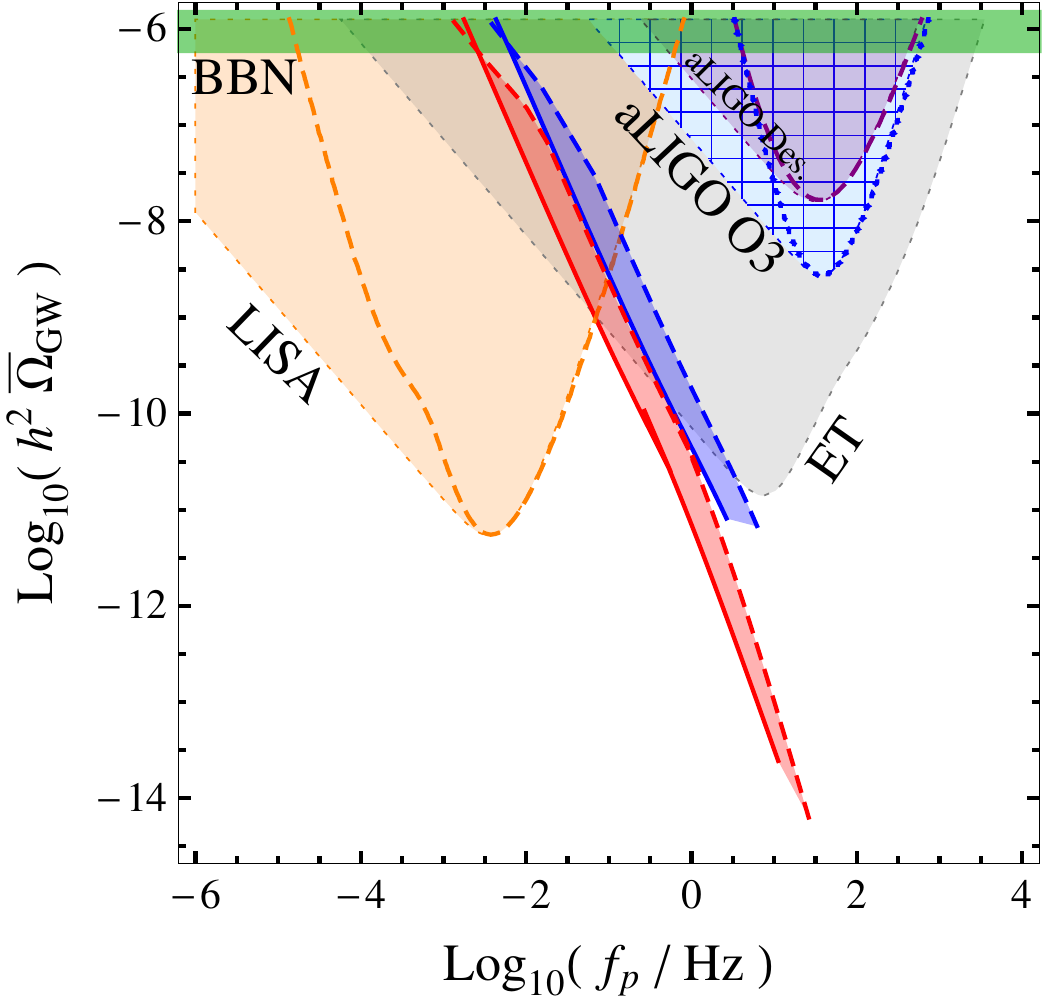} \\
\end{tabular}
 \caption{\small \it Experimental parameter reach in the $\{ h^2 \overline\Omega_{\textrm{\tiny GW}},f_p\}$ plane for SGWBs in the regimes $\Omega_{\textrm{\tiny GW}}^{\rm env}$ (regions inside dotted borders) and $\Omega_{\textrm{\tiny GW}}^{\rm sw}$ (regions inside dashed borders). Diagonal strips are for $N=10$ (red), $25$ (blue) for $\rho = 1$\,TeV (left panels),  $\rho = 100$\,TeV (right panels), SNR $\ge 2$ (upper panels) and SNR $\ge 1000$ (lower panels). Solid and dashed lines on the edge of the strips correspond to the regime $\overline\Omega_{\textrm{\tiny GW}}\simeq \overline\Omega_{\textrm{\tiny GW}}^{\rm env}$ and $\overline\Omega_{\textrm{\tiny GW}}^{\rm sw}$, respectively. Shadowed (green) regions in both panels are excluded by BBN. We have considered for the wall velocity $v_\omega\simeq 0.99$.}
\label{fig:Omega_fp}
\end{figure*}

In Fig.~\ref{fig:alpha_beta} we show the curves of $\alpha$ (left panel) and $\beta/H_\star$ (right panel) as functions of $\lambda_1$, for some values of $N$ and $\rho$. The value of $h^2 \overline \Omega_\GW^{\env (\sw)}$ and $f_p$ obtained along these curves are presented in Fig.~\ref{fig:Omega_fp_TR}. Here $h\simeq 0.67$ is the scaling factor for the  Hubble parameter and is introduced to keep the result independent of the precise value of today's Hubble parameter.
In Fig.~\ref{fig:Omega_fp} we display the detectability prospects to the parameter space $\{ h^2 \overline\Omega_{\GW},f_p\}$. Within each detector region, the SGWB sourced by a first-order phase transition following the envelope (regions inside the dotted borders) and sound wave (regions inside the dashed borders) behavior, is detected with signal-to-noise ratio SNR$\,\ge 2$ (upper panels) and  SNR$\,\ge 1000$ (lower panels). The SNR is calculated as
\begin{equation}
{\rm SNR} = \sqrt{(3.16\times 10^7 s)\frac{\mathcal T}{\textrm{1 year}}  \int_0^\infty df \frac{\Omega_{\rm GW}^2(f)}{\Omega_{\rm sens}^2(f)}     }~,
\end{equation}
where $\Omega_{\rm sens}(f)$ is the sensitivity curve of the experiment and  $\mathcal T$ is the total time (in years) during which a given experiment takes data. For concreteness, we have used 3 years for LISA, 7 years for ET, and 8 years for aLIGO design. The parameter region labeled BBN and aLIGO O3 are in tension with data~\cite{Megias:2020vek,Abbott:2021xxi}~\footnote{Ref.~\cite{Megias:2020vek} determined the equivalent constraint due to O2-run data. Here we follow the same procedure but with the up-to-date bound~\cite{Abbott:2021xxi}.}. The fact that in the lower panels the aLIGO design detection region falls inside the LIGO O3 exclusion region indicates that no first order phase transition with SNR larger than 1000 will be ever observed with the aLIGO design sensitivity since such a powerful signal is already ruled out.

In Fig.~\ref{fig:Omega_fp} the diagonal strips correspond to the predictions in our warped setup for $\rho = 1\textrm{ TeV}$ (left panels) and $\rho = 100\textrm{ TeV}$ (right panels). Solid and dashed lines on the edge of the strips are evaluated in the approximation   $\overline\Omega_{\textrm{\tiny GW}}\simeq \overline\Omega_{\textrm{\tiny GW}}^{\rm env}$ and $\overline\Omega_{\textrm{\tiny GW}}\simeq \overline\Omega_{\textrm{\tiny GW}}^{\rm sw}$.
We also analyze  values of $\rho>100$ TeV which do not appear in Fig.~\ref{fig:Omega_fp}. Taking into account that $m_\KK/\rho\simeq 3.88$ as discussed in Sec.~\ref{subsec:radion},  we find that values of $m_{\KK} \sim \mathcal{O}(10^{4})\,$ -- $\mathcal{O}(10^{7})\,$TeV are being already cornered by current aLIGO O3 data. The forecast shows that by the early 2040s the ground-based and space-based interferometer network will be sensitive to a huge parameter region, reaching the following KK scales: $m_\KK \lesssim 10^5$~TeV in LISA, $10^2~\TTeV \lesssim m_\KK \lesssim 10^8$~TeV in LIGO-like detectors, and $m_\KK \lesssim 10^9$~TeV in ET. Most of the SGWB signals are so inside the sensitivity curves that they are expected to be reconstructed with a very high accuracy, shedding light on the scale of new physics~\cite{Figueroa:2018xtu, Caprini:2019pxz, Flauger:2020qyi}.

\section{Conclusions}
\label{sec:conclusions}

We have studied warped models with a radion stabilized with a polynomial bulk potential, and addressed the computation of the radion effective potential at zero and finite temperature.  The zero temperature computation has been carried out by adopting a recent technique that allows for a sizable backreaction on the gravitational metric. In the limit of small backreaction, our findings are in good agreement with previous results in the literature, as for instance the original Goldberger-Wise potential~\cite{Goldberger:1999uk}.
At finite temperature, the radion undergoes a first order phase transition which, in the four-dimensional language, can be seen as a  confinement/deconfinement phase transition.
We have found that the radion phase transition potentially generates a SGWB signal which covers several gravitational wave frequency bands. In particular, it can be often detected at both ground-based and space-based gravitational wave interferometers with a high signal-to-noise ratio, enabling an accurate signal reconstruction at completely different frequency ranges. In view of the correct LHC constraints,  we have considered the possibility of Kaluza-Klein resonance masses much heavier than the TeV scale, thus leaving open the door for a little hierarchy problem and the corresponding level of fine-tuning. By recasting the sensitivity region of the gravitational wave detector network, we have found that in the next two decades the gravitational wave detectors will broadly probe warped models with resonances up to $m_\KK \lesssim 10^9 \, \TTeV$. However, the most interesting detection prospects seem to be those for the region $1 \, \TTeV \ll m_\KK \lesssim 100 \, \TTeV$, which deserves a future, dedicated study. In fact, for this region, the synergy between  gravitational wave and collider experiments is exciting.

%%%%%%%%%%%%%%%%%%%%%%%%%%%
% Acknowledgments
\ack 
%%%%%%%%%%%%%%%%%%%%
We would like to thank Huaike Guo for discussions. The work of EM is
supported by the Spanish MINEICO under Grant FIS2017-85053-C2-1-P, by
the FEDER/Junta de Andaluc\'{\i}a-Consejer\'{\i}a de Econom\'{\i}a y
Conocimiento 2014-2020 Operational Programme under Grant
A-FQM-178-UGR18, by Junta de Andaluc\'{\i}a under Grant FQM-225, and
by Consejer\'{\i}a de Conocimiento, Investigaci\'on y Universidad of
the Junta de Andaluc\'{\i}a and European Regional Development Fund
(ERDF) under Grant SOMM17/6105/UGR. The research of EM is also
supported by the Ram\'on y Cajal Program of the Spanish MINEICO under
Grant RYC-2016-20678. GN is partly supported by the ROMFORSK grant
Project.~No.~302640 ``Gravitational Wave Signals From Early Universe
Phase Transitions''. The work of MQ is partly supported by the
Secretaria d'Universitats i Recerca del Departament d'Empresa i
Coneixement de la Generalitat de Catalunya under the grant
2017SGR1069, by the Ministerio de Econom\'{i}a, Industria y
Competitividad under grant FPA2017-88915-P, and from the Centro de
Excelencia Severo Ochoa under the grant SEV-2016-0588.  IFAE is
partially funded by the CERCA program of the Generalitat de Catalunya.

%\vspace*{-1cm}
%\vfill \eject
%%%%%%%%%%%

%%%%%%%%%%%
\section*{References}
%\input{refs}
%\bibliographystyle{elsarticle-num}
%\bibliography{refs}

\begin{thebibliography}{10}
\expandafter\ifx\csname url\endcsname\relax
  \def\url#1{\texttt{#1}}\fi
\expandafter\ifx\csname urlprefix\endcsname\relax\def\urlprefix{URL }\fi
\expandafter\ifx\csname href\endcsname\relax
  \def\href#1#2{#2} \def\path#1{#1}\fi

\bibitem{Randall:1999ee}
L.~Randall, R.~Sundrum, {A Large mass hierarchy from a small extra dimension},
  Phys. Rev. Lett. 83 (1999) 3370--3373.

\bibitem{Sirunyan:2018ryr}
A.~M. Sirunyan, et~al., {Search for resonant $ \mathrm{t}\overline{\mathrm{t}}
  $ production in proton-proton collisions at $ \sqrt{s}=13 $ TeV}, JHEP 04
  (2019) 031.

\bibitem{Aaboud:2019roo}
M.~Aaboud, et~al., {Search for heavy particles decaying into a top-quark pair
  in the fully hadronic final state in $pp$ collisions at $\sqrt{s} =$ 13 TeV
  with the ATLAS detector}, Phys. Rev. D99~(9) (2019) 092004.

\bibitem{Creminelli:2001th}
P.~Creminelli, A.~Nicolis, R.~Rattazzi, {Holography and the electroweak phase
  transition}, JHEP 03 (2002) 051.

\bibitem{Randall:2006py}
L.~Randall, G.~Servant, {Gravitational waves from warped spacetime}, JHEP 05
  (2007) 054.

\bibitem{Nardini:2007me}
G.~Nardini, M.~Quiros, A.~Wulzer, {A Confining Strong First-Order Electroweak
  Phase Transition}, JHEP 09 (2007) 077.

\bibitem{Caprini:2015zlo}
C.~Caprini, et~al., {Science with the space-based interferometer eLISA. II:
  Gravitational waves from cosmological phase transitions}, JCAP 1604~(04)
  (2016) 001.

\bibitem{Goldberger:1999wh}
W.~D. Goldberger, M.~B. Wise, {Bulk fields in the Randall-Sundrum
  compactification scenario}, Phys. Rev. D60 (1999) 107505.

\bibitem{Csaki:2000zn}
C.~Csaki, M.~L. Graesser, G.~D. Kribs, {Radion dynamics and electroweak
  physics}, Phys. Rev. D63 (2001) 065002.

\bibitem{Cabrer:2011fb}
J.~A. Cabrer, G.~von Gersdorff, M.~Quiros, {Suppressing Electroweak Precision
  Observables in 5D Warped Models}, JHEP 05 (2011) 083.

\bibitem{Megias:2018sxv}
E.~Megias, G.~Nardini, M.~Quiros, {Cosmological Phase Transitions in Warped
  Space: Gravitational Waves and Collider Signatures}, JHEP 09 (2018) 095.

\bibitem{Carena:2018cow}
M.~Carena, E.~Megias, M.~Quiros, C.~Wagner, {$ {R}_{D^{\left(*\right)}} $ in
  custodial warped space}, JHEP 12 (2018) 043.

\bibitem{Megias:2020vek}
E.~Megias, G.~Nardini, M.~Quiros, {Gravitational Imprints from Heavy
  Kaluza-Klein Resonances}, Phys. Rev. D102~(5) (2020) 055004.

\bibitem{Goldberger:1999uk}
W.~D. Goldberger, M.~B. Wise, {Modulus stabilization with bulk fields}, Phys.
  Rev. Lett. 83 (1999) 4922--4925.

\bibitem{Papadimitriou:2007sj}
I.~Papadimitriou, {Multi-Trace Deformations in AdS/CFT: Exploring the Vacuum
  Structure of the Deformed CFT}, JHEP 05 (2007) 075.

\bibitem{Megias:2014iwa}
E.~Megias, O.~Pujolas, {Naturally light dilatons from nearly marginal
  deformations}, JHEP 08 (2014) 081.

\bibitem{Lizana:2019ath}
J.~M. Lizana, M.~Olechowski, S.~Pokorski, {A new way of calculating the
  effective potential for a light radion}, JHEP 09 (2020) 092.

\bibitem{Megias:2015ory}
E.~Megias, O.~Pujolas, M.~Quiros, {On dilatons and the LHC diphoton excess},
  JHEP 05 (2016) 137.

\bibitem{Coleman:1977py}
S.~R. Coleman, {The Fate of the False Vacuum. 1. Semiclassical Theory}, Phys.
  Rev. D15 (1977) 2929--2936, [erratum: Phys. Rev. D16, 1248(1977)].

\bibitem{Linde:1980tt}
A.~D. Linde, {Fate of the False Vacuum at Finite Temperature: Theory and
  Applications}, Phys. Lett. 100B (1981) 37--40.

\bibitem{Konstandin:2010cd}
T.~Konstandin, G.~Nardini, M.~Quiros, {Gravitational Backreaction Effects on
  the Holographic Phase Transition}, Phys. Rev. D82 (2010) 083513.

\bibitem{Bunk:2017fic}
D.~Bunk, J.~Hubisz, B.~Jain, {A Perturbative RS I Cosmological Phase
  Transition}, Eur. Phys. J. C78~(1) (2018) 78.

\bibitem{Huber:2008hg}
S.~J. Huber, T.~Konstandin, {Gravitational Wave Production by Collisions: More
  Bubbles}, JCAP 0809 (2008) 022.

\bibitem{Hindmarsh:2017gnf}
M.~Hindmarsh, S.~J. Huber, K.~Rummukainen, D.~J. Weir, {Shape of the acoustic
  gravitational wave power spectrum from a first order phase transition}, Phys.
  Rev. D96~(10) (2017) 103520, [erratum: Phys. Rev. D101, no.8, 089902(2020)].

\bibitem{Caprini:2019egz}
C.~Caprini, et~al., {Detecting gravitational waves from cosmological phase
  transitions with LISA: an update}, JCAP 03 (2020) 024.

\bibitem{Guo:2020grp}
H.-K. Guo, K.~Sinha, D.~Vagie, G.~White, {Phase Transitions in an Expanding
  Universe: Stochastic Gravitational Waves in Standard and Non-Standard
  Histories}, JCAP 2101 (2021) 001.

\bibitem{Hindmarsh:2020hop}
M.~B. Hindmarsh, M.~L\"uben, J.~Lumma, M.~Pauly, {Phase transitions in the early
  universe}, SciPost Phys. Lect. Notes 24 (2021) 1.

\bibitem{Ares:2020lbt}
F.~R. Ares, M.~Hindmarsh, C.~Hoyos, N.~Jokela, {Gravitational waves from a
  holographic phase transition}, arXiv:2011.12878.

\bibitem{Abbott:2021xxi} R.~Abbott, et~al., {Upper Limits on the
  Isotropic Gravitational-Wave Background from Advanced LIGO's and
  Advanced Virgo's Third Observing Run}, arXiv:2101.12130.

\bibitem{Figueroa:2018xtu}
D.~G. Figueroa, E.~Megias, G.~Nardini, M.~Pieroni, M.~Quiros, A.~Ricciardone,
  G.~Tasinato, {LISA as a probe for particle physics: electroweak scale tests
  in synergy with ground-based experiments}, PoS GRASS2018 (2018) 036.

\bibitem{Caprini:2019pxz}
C.~Caprini, D.~G. Figueroa, R.~Flauger, G.~Nardini, M.~Peloso, M.~Pieroni,
  A.~Ricciardone, G.~Tasinato, {Reconstructing the spectral shape of a
  stochastic gravitational wave background with LISA}, JCAP 11 (2019) 017.

\bibitem{Flauger:2020qyi}
R.~Flauger, N.~Karnesis, G.~Nardini, M.~Pieroni, A.~Ricciardone, J.~Torrado,
  {Improved reconstruction of a stochastic gravitational wave background with
  LISA}, JCAP 01 (2021) 059.

\end{thebibliography}
%%%%%%%%%%%

\end{document}